%% file: main.tex
\documentclass[conference,10pt]{IEEEtran}
\IEEEoverridecommandlockouts

\usepackage{cite}

\usepackage[pdftex]{graphicx}
\usepackage{cite}

\usepackage{amsmath}
\usepackage{algorithm}
\usepackage{algorithmic}
\usepackage{comment}
\usepackage{color}
\usepackage{amsmath}
\usepackage{amsfonts, amssymb, amsbsy}

\usepackage{mathtools}

\usepackage{amsthm}

\input{defines2}

\newcommand{\bPhi}{\mathbf{\Phi}}

\newcommand{\bmhh}{\widehat{\bmh}}

\newcommand{\vect}[1]{\mathbf{#1}}
\def\Htran{\mbox{\tiny $\mathrm{H}$}}
\def\Ttran{\mbox{\tiny $\mathrm{T}$}}
\newcommand{\mathacr}[1]{\mathsf{#1}}

\makeatletter
\newcommand\fs@spaceruled{\def\@fs@cfont{\bfseries}\let\@fs@capt\floatc@ruled
  \def\@fs@pre{\vspace{0.5\baselineskip}\hrule height.8pt depth0pt \kern2pt}%
  \def\@fs@post{\kern1pt\hrule\relax}%
  \def\@fs@mid{\kern2pt\hrule\kern2pt}%
  \let\@fs@iftopcapt\iftrue}
\makeatother

\begin{document}

\title{A Rate Splitting Strategy for Mitigating Intra-Cell Pilot Contamination in Massive MIMO\vspace{-0.3cm} \\ \thanks{\scriptsize EURECOM's research is partially supported by its industrial members:
ORANGE, BMW, Sy\-man\-tec, SAP, Monaco Telecom, iABG,  and by the projects DUPLEX (French ANR), MASS-START (French FUI) and EU ITN project SPOTLIGHT. L. Sanguinetti was supported by the University of Pisa under the PRA 2018-2019 Research Project CONCEPT and by the Italian Ministry of Education and Research (MIUR) in the framework of the CrossLab project (Departments of Excellence). }}

\author{
\IEEEauthorblockN{
Christo Kurisummoottil Thomas\IEEEauthorrefmark{3}, Bruno Clerckx\IEEEauthorrefmark{2}, Luca Sanguinetti\IEEEauthorrefmark{5}, Dirk Slock \IEEEauthorrefmark{3}
\vspace{1mm}}
\IEEEauthorblockN{
\IEEEauthorrefmark{3}EURECOM, Sophia-Antipolis, France, Email: \{kurisumm,slock\}@eurecom.fr\\
\IEEEauthorrefmark{2}Imperial College London, Email: b.clerckx@imperial.ac.uk \\
\IEEEauthorrefmark{5}University of Pisa, Italy, Email: luca.sanguinetti@unipi.it \vspace{-0.6cm}
}}

\maketitle
\begin{abstract}
The spectral efficiency (SE) of Massive MIMO (MaMIMO) systems is affected by low quality channel estimates. 
Rate-Splitting (RS) has recently gained some interest in multi-user multiple antenna systems as an effective means to mitigate the multi-user interference due to imperfect channel state information. This paper investigates the benefits of RS in the downlink of a single-cell MaMIMO system when all the users use the same pilot sequence for channel estimation. Novel  expressions  for  the SE achieved in the downlink by a single-layer RS strategy (that relies on a single successive interference cancellation at each user side) are derived and used to design precoding schemes and power allocation strategies for common and private messages. Numerical results are used to show that the proposed RS solution achieves higher spectral efficiency that conventional MaMIMO with maximum ratio precoding. 
\end{abstract}
\vspace{-0.3cm}
\section{Introduction}

Massive MIMO (MaMIMO) is a wireless technology where the base stations (BSs) are equipped with a large number $M$ of antennas to serve a multitude of single-antenna $K$ user equipments (UEs) by spatial multiplexing \cite{MarzettaTWC2010}. The acquisition of channel state information (CSI) is the limiting factor in MaMIMO \cite{MarzettaTWC2010}. In a time-division duplex (TDD) mode, channel reciprocity allows to acquire all the necessary CSI for uplink (UL) and downlink (DL) transmissions from a finite number of UL pilot signals \cite{MarzettaTWC2010}. Thanks to the intense research performed over the last decade, 
MaMIMO is today a mature technology \cite{MarzettaBook,massivemimobook}, which has been adopted into the 5G NR standard \cite{FiveResearchDirections}. 

One phenomenon that is tightly connected with MaMIMO is \emph{pilot contamination}, which can be briefly explained as follows \cite{MarzettaTWC2010}. UEs that transmit the same pilot signal contaminate each others' channel estimates. This ''pilot interference'' not only reduces the CSI quality but also creates the so-called ''coherent interference'', which has been believed to fundamentally limit the spectral efficiency (SE) of MaMIMO, even when $M\to \infty$ \cite{MarzettaTWC2010,MarzettaBook}. Recently, \cite{BjornsonTWC2018} showed that with optimal signal processing and spatially correlated channels, the SE increases without bound as $M\to \infty$ while $K$ is fixed. The fact that there is no fundamental SE limit does not imply that the pilot contamination effect disappears; there is still an SE loss caused by estimation errors and interference rejection \cite{Sanguinetti2019a}. The aim of this paper is to deal with this effect for a finite $M$.

Observe that, when the estimation error variance decays with the signal-to-noise-ratio (SNR) as $\mathcal{O}({\rm{SNR}}^{-\delta})$ for some $0 \leq \delta < 1$, conventional precoding techniques result in a sum degrees of freedom (DoF) of $K\delta$.  This in turn reveals that as $\delta \rightarrow 0$ (implies constant channel estimation error), the system becomes interference limited. A possible solution to this issue is to take a rate splitting (RS) approach \cite{ClerckxCommMag2016} that splits the UEs' messages into common and private parts, encode the commmon parts into a common stream, and private
parts into private streams and superpose in a non-orthogonal manner the common stream on top
of all private streams. The common stream is drawn from a codebook shared by all UEs
and is intended to one only, but is decodable by all UEs. On the other hand, the private streams are to be decoded
by the corresponding UEs only. The sum DoF achieved by RS in the DL is $1+(K-1)\delta$ \cite{HamdiTCOM2016}, which is higher than $K\delta$ and matches the upper bound
obtained from the Aligned Image Sets in \cite{DavoodiTIT2016}. Interestingly, RS not only achieves the optimal sum-DoF but the entire DoF region of the $K$-UE channel with imperfect CSI \cite{PiovanoCL2017}. 

Motivated by the above results, the design and optimization of RS at finite values of SNR has been investigated and was found to provide significant benefits in the DL with imperfect CSI, compared to multi-user MIMO and NOMA \cite{HamdiTCOM2016,JoudehTSP2016,MaoClerckxJWCN2018}, but also to Dirty Paper Coding \cite{MaoClerckxArxiv019}. The application of RS to an FDD MaMIMO system has been investigated in \cite{MDaiTWC2016,PapazafeiropoulosTVT2017}. Particularly, \cite{MDaiTWC2016} shows that a two-layer RS architecture, so-called hierarchical RS (HRS), can bring significant benefits in MaMIMO.
 
In this paper, we focus on a TDD single-cell MaMIMO network and assume that all the UEs use the same pilot signal for channel estimation. Novel expressions for the SE achieved in the DL by a single-layer RS strategy are derived by applying the hardening bound to both common and private messages \cite{MarzettaBook}. A maximum ratio (MR) precoding scheme is used for private streams while a precoder based on a weighted combination of the channel estimates of all UEs is adopted for the common stream. A novel algorithm is proposed to allocate the power among the common and private streams. 

\section{System model}
We consider a single-cell MaMIMO network where the BS is equipped with $M$ antennas and serves $K$ UEs. We denote $\vect{h}_{i}\in \mathbb{C}^{M}$ the channel from UE $i$ to the BS, and consider a correlated Rayleigh block fading model $\vect{h}_{i} \sim \CN (\vect{0}, \vect{R}_{i})$ where $\vect{R}_{i} \in \mathbb{C}^{M \times M}$ is the  covariance matrix \cite[Sec. 2.2]{massivemimobook}. The Gaussian distribution is used to model the small-scale fading variations, while $\vect{R}_{i}$ describes the macroscopic propagation characteristics. The normalized trace $\beta_{i} = \frac{1}{M} \tr(\vect{R}_{i})$ is the average channel gain from the BS to UE $i$.

The UEs are perfectly synchronized and operate according to a TDD protocol with a data transmission phase and a pilot phase for channel estimation \cite{massivemimobook}. We consider the standard block fading TDD protocol in which each coherence block consists of $\tau$ channel uses, whereof $\tau_p$ are used for UL pilots, $\tau_u$ for UL data, and $\tau_d$ for DL data, with $\tau = \tau_p + \tau_u + \tau_d$. Only the DL is considered in this paper, i.e., $\tau_u =0$.
\vspace{-0.3cm}
\subsection{Channel estimation}
We assume that a single pilot sequence of length $\tau_p$ is used. For a total uplink pilot power of $\rho_{\rm{tr}}$ per UE, the BS obtains the MMSE estimate of $\vect{h}_{i}$ as
\begin{align}
\!\!\!\!\widehat{\vect{h}}_{i} = \vect{R}_{i} {\vect{Q}} ^{-1} \bigg( \sum_{k =1}^K \vect{h}_{k} + \frac{1}{\sqrt{\rho_{\rm{tr}}}} \vect{n}_{i}   \bigg) \!\sim \!\CN \left( \vect{0},  \vect{\Phi}_{i} \right)
\end{align}
where $\vect{n}_{i} \sim \CN (\vect{0}, \vect{I}_{M})$ is noise, $
\vect{\Phi}_{i}  = \vect{R}_{i} \vect{Q} ^{-1} \vect{R}_{i}$ and $\vect{Q} = \sum_{k =1}^K \vect{R}_{k} + \frac{1}{\rho_{\rm{tr}}} \vect{I}_{M}$.
The estimation error $\widetilde{\vect{h}}_{i} = \vect{h}_{i} - \widehat{\vect{h}}_{i}  \sim \CN \left( \vect{0}, \vect{R}_{i}- \vect{\Phi}_{i} \right)$ is independent of $\widehat{\vect{h}}_{i}$. The mutual interference generated by the pilot-sharing UEs is known as pilot contamination and has two main consequences in the channel estimation process. The first is the reduced estimation quality, whereas the second is that the estimates $\{\widehat{\vect{h}}_{i}\}$ become correlated. If $\vect{R}_{k}$ is invertible, we have that \cite[Sec. 3.2]{massivemimobook}
\begin{align}\label{eq:correlated_channel_estimates}
\widehat{\vect{h}}_{i}   = \vect{R}_{i}\vect{R}_{k}^{-1}\widehat{\vect{h}}_{k} 
\end{align}
from which it follows that $\mathbb{E}\{\widehat{\vect{h}}_{i} \widehat{\vect{h}}_{k}^{\Htran} \} = \vect{R}_{i} \vect{Q} ^{-1} \vect{R}_{k}$.

\begin{figure*}
\begin{align}\tag{11}
{\gamma}_{k,c} =  \frac{ \rho_c| \mathbb{E}\{ \vect{h}_{k}^{\Htran}{\vect{w}}_{c} \} |^2  }{ \sum\limits_{i=1}^{K}  {\rho_i}\mathbb{E} \{  | \vect{h}_{k}^{\Htran}\vect{w}_{i}   |^2 \} + 
\rho_c \Big(\mathbb{E} \{  | \vect{h}_{k}^{\Htran}  {\vect{w}}_{c} |^2 \} - | \mathbb{E}\{\vect{h}_{k}^{\Htran} {\vect{w}}_{c}  \} |^2\Big) + \sigma^2} \label{eq:gamma_c}
\end{align}
\hrule
\begin{align}\tag{12}
\!\!\!\gamma_{k} =  \frac{ {\rho_k}| \mathbb{E}\{ \vect{h}_{k}^{\Htran} \vect{w}_{k} \} |^2  }{ \sum\limits_{i=1}^{K}  {\rho_i} \mathbb{E} \{  | \vect{h}_{k}^{\Htran} \vect{w}_{i}  |^2 \} 
- {\rho_k}| \mathbb{E}\{ \vect{h}_{k}^{\Htran}\vect{w}_{k}  \} |^2 + \rho_c \Big(\mathbb{E} \{  | \vect{h}_{k}^{\Htran} {\vect{w}}_{c}  |^2 \} - | \mathbb{E}\{\vect{h}_{k}^{\Htran} {\vect{w}}_{c} \} |^2\Big) + \sigma^2}\label{eq:gamma_p}
\end{align}
\hrule
\end{figure*}
\vspace{-0.3cm}

\subsection{Rate Splitting in Downlink transmissions}

The RS scheme is used in the DL for transmission. The message intended to UE $k$ is split into two parts, $W_k = (W_{k0},\,W_{k1})$. We assume that $W_{k0} \in \mathcal{W}_{k0}$ represents the common part and $W_{k1} \in \mathcal{W}_{k1}$ is the private part. All the common parts are packed into one common message, $W_c = (W_{k0},\,...,\,W_{K0}) \in \mathcal{W}_c$, which is encoded into a common stream ${\varsigma}_c$ using a common codebook. The private message $W_{k1}$ is encoded in the conventional manner into the private stream ${\varsigma}_k$.  The resulting transmitted DL signal is:
\begin{align}
\vect{x} = \underbrace{{\vect{w}}_c {\varsigma}_c}_{\text{Common message}} + \sum_{i=1}^{K} \underbrace{\vect{w}_{i} \varsigma_{i}} _{\text{Private messages}}
\end{align}
where  $\varsigma_{i} \sim \CN(0,\rho_i)$ is assigned to a precoding vector $ \vect{w}_{i} \in \mathbb{C}^{M}$ that determines the spatial directivity of the transmission and satisfies $ \mathbb{E}\{\|  \vect{w}_{i} \|^2\}    =1$ so that $\rho_i$ represents the average transmit power of UE $\forall i$. Similarly, ${\varsigma}_c\sim \CN(0,\rho_c)$ denotes the common message, which is assigned to a precoding vector $ \vect{w}_{c} \in \mathbb{C}^{M}$ with $\mathbb{E}\{\|  \vect{w}_{c} \|^2\}   =1$ so that $\rho_c$ represents its average transmit power. We assume that \begin{align}
\rho_c + \sum\limits_{i=1}^K\rho_i \le \rho_T
\label{eq_power_constraints}
\end{align}
where $\rho_T$ is the total transmit power in the DL. The received signal $y_{k} \in \mathbb{C}$ at UE~$k$ is given by
\begin{align}
y_{k} &= \vect{h}_{k}^{\Htran}{\vect{w}}_c {\varsigma}_c +  \vect{h}_{k}^{\Htran} \vect{w}_{k}\varsigma_{k} + \sum_{i=1, i\ne k}^{K}  \vect{h}_{k}^{\Htran} \vect{w}_{i}\varsigma_{i}  + n_{k}\label{eq:downlink-signal-model}
\end{align}
where $n_{k} \sim \CN(0,\sigma^2)$ is the receiver noise. At each UE $k$, the common stream is first decoded into $\widehat{W}_c$, by treating the interference from the private streams as noise. Then, successive interference cancellation (SIC) is performed, which removes the common message part from the received signal. Further, the private stream ${\varsigma}_k$ is decoded into $\widehat{W}_{k1}$ by treating the intra-cell interference as noise. UE $k$ reconstructs the transmitted message by extracting $\widehat{W}_{k0}$ from $\widehat{W}_{c}$. Further, combining with the decoded private stream to form $\widehat{W}_k = (\widehat{W}_{k0},\widehat{W}_{k1})$.

\subsection{Spectral efficiency}

Characterizing the SE in the DL is hard since it is unclear how UE $k$ should best estimate the effective precoded channels $ \vect{h}_{k}^{\Htran} {\vect{w}}_{c}$ and $ \vect{h}_{k}^{\Htran} \vect{w}_{k}$ that are needed for decoding the common signal ${\varsigma}_{c}$ and the private signal ${\varsigma}_{k}$. A common approach in classical MaMIMO is to resort to the \emph{hardening bound} \cite[Sec. 4.3]{massivemimobook}. This bound relies on the assumption that the deterministic average precoded channels $\mathbb{E}\{\vect{h}_{k}^{\Htran}{\vect{w}}_c\}$ and $\mathbb{E}\{\vect{h}_{k}^{\Htran} \vect{w}_{k}\}$ are known at UE $k$. The received signals for the common and private messages can then be expressed as
\begin{align}\notag
y_{k,c} = \mathbb{E}\{\vect{h}_{k}^{\Htran}{\vect{w}}_c\} {\varsigma}_c &+ \left(\vect{h}_{k}^{\Htran}{\vect{w}}_c- \mathbb{E}\{\vect{h}_{k}^{\Htran}{\vect{w}}_c\}\right){\varsigma}_c \\ & + \sum_{i=1}^{K}  \vect{h}_{k}^{\Htran} \vect{w}_{i}\varsigma_{i}  + n_{k}\label{eq:downlink-signal-model-common}
\end{align}
and (after SIC)
\begin{align}\notag
y_{k,p}&= \mathbb{E}\{\vect{h}_{k}^{\Htran}{\vect{w}}_k\} {\varsigma}_k + \left(\vect{h}_{k}^{\Htran}{\vect{w}}_k - \mathbb{E}\{\vect{h}_{k}^{\Htran}{\vect{w}}_k\}\right){\varsigma}_k \\&+  \left(\vect{h}_{k}^{\Htran}{\vect{w}}_c - \mathbb{E}\{\vect{h}_{k}^{\Htran}{\vect{w}}_c\}\right){\varsigma}_c + \sum_{i=1, i\ne k}^{K}  \vect{h}_{k}^{\Htran} \vect{w}_{i}\varsigma_{i}  + n_{k}.\label{eq:downlink-signal-model-private}
\end{align}
The following bounds can be computed.
\begin{lemma}\label{Theorem1}
Achievable rates for the common and private messages of UE $k$ can be computed as 
\begin{equation} \label{eq:downlink-SE-expression-forgetbound}
\mathacr{SE}_{k,c} = \frac{\tau_d}{\tau} \log_2   (  1 +
{\gamma}_{k,c} )
\end{equation}
and
\begin{equation} \label{eq:downlink-SE-expression-forgetbound}
\mathacr{SE}_{k}  = \frac{\tau_d}{\tau} \log_2   (  1 +
\gamma_{k} )
\end{equation}
with ${\gamma}_{k,c}$ and $\gamma_{k} $ given by \eqref{eq:gamma_c} and \eqref{eq:gamma_p}. The expectations are computed over channel realizations.
\end{lemma}
\begin{IEEEproof}
\setcounter{equation}{12}
It can be proved from \eqref{eq:downlink-signal-model-common} and \eqref{eq:downlink-signal-model-private} by using standard results in MaMIMO (e.g., \cite[App. C.3.6]{massivemimobook}), which are omitted for space limitations.
\end{IEEEproof}
The achievable rate of the common message is defined as 
\begin{equation}\label{eq_min_SE}
\mathacr{SE}_{c} = \frac{\tau_d}{\tau}\log(1+{\gamma}_{l_{\min},c})
\end{equation}
where 
\beq
\begin{array}{l}
l_{\min} = \arg \min_{k} \gamma_{k,c}
\label{eq_min_gamma_c}
 \end{array}
 \eeq
Observe that the above achievable rates can be utilized along with any precoding scheme. Moreover, each of the expectations in ${\gamma}_{k,c}$ and $\gamma_{k} $ can be computed separately by means of Monte Carlo simulations. Closed forms will be provided next for the proposed precoding schemes.

\section{Power optimization and precoding design} \label{Section_POpt}
A common and popular choice for ${\bf{w}}_k$ is MR precoding, defined as
\begin{align}
\bmw_k^{\mathsf{MR}} = \frac{\bmhh_{k}}{\sqrt{\E\{|\bmhh_{k}|^2\}}} = \frac{\bmhh_{k}}{\sqrt{\tr\{\bPhi_{k}\}}}
\end{align} 
which has low computational complexity and allows to compute some of the expectations in closed form. Particularly, we have that (e.g., \cite[App. C.3.7]{massivemimobook})
\begin{align}\label{eq:MR_expectation1}
|\mathbb{E}\{ \vect{h}_{k}^{\Htran} \bmw_k^{\mathsf{MR}} \}|^2 &= {\tr\{\bPhi_{k}\}}
\\\label{eq:MR_expectation2}
\mathbb{E} \{  | \vect{h}_{k}^{\Htran} \vect{w}_{i}^{\mathsf{MR}}  |^2 \} &=  \frac{\tr\{\bmR_{k}\bPhi_{i}\} + \Big|\tr\{\bmR_{k}\vect{Q}^{-1}\vect{R}_i\}\Big|^2}{\tr\{\bPhi_{i}\}}.
\end{align} 
In the remainder, we assume that MR precoding is used for private messages. 
Next, we look for the transmit powers that maximize the sum SE of the network and design the precoding vector for the common message. 

\subsection{Power optimization}
From the above section, the sum SE, for any given precoding scheme, can be computed as:
\vspace{-1mm}\begin{align}\label{eq:sum_SE}
\mathacr{SE} = \mathacr{SE}_c + \sum_{k=1}^K\mathacr{SE}_k
\end{align}
where $\mathacr{SE}_k$ and $\mathacr{SE}_c$ are given in \eqref{eq:downlink-SE-expression-forgetbound} and \eqref{eq_min_SE}, respectively.
The power allocation problem can thus be formulated as:
\vspace{-1mm}\begin{align}
\max\limits_{\{\rho_c\ge 0,\boldsymbol{\rho}\ge {\bf 0}\}} \quad &   \mathacr{SE}_c (\rho_c,\boldsymbol{\rho}) + \sum_{k=1}^K\mathacr{SE}_k(\rho_c,\boldsymbol{\rho})\\ \mbox{s.t.} \quad \quad \;\,&  \rho_c + \sum\limits_{i=1}^K\rho_i \le \rho_T
\label{eq_wc_opt_max_sum_SE}
\vspace{-1mm}\end{align}
with $\boldsymbol{\rho} = [\rho_1,\ldots,\rho_K]^{\Ttran}$. Finding the solution to the above problem is a challenge since it is not in a convex form. A possible way out consists in using the method in \cite{KimGiannakisAllerton2008}, and linearize the sum SE in \eqref{eq:sum_SE} using a first order Taylor series approximation. The optizimation is then carried out by adopting an iterative approach in which the variables $\rho_c$ and $\{\rho_i : i = 1,\ldots, K\}$ are alternatively optimized. In Appendix A, it is shown that at iteration $t$ the powers must be updated as follows
\vspace{-1mm}\beq\label{eq:optimal_k}
\rho_k^{(t)} = \left(\frac{1}{\mu^{(t)}+\sigma_{k}^{(2,t)}}-\frac{1}{\sigma_{k}^{(1,t)}}\right)^{+} 
\eeq
and
\beq\label{eq:optimal_c}
\rho_{c}^{(t)} = \left(\frac{1}{\mu^{(t)}+\sigma_{c}^{(2,t)}}-\frac{1}{\sigma_{c}^{(1,t)}}\right)^{+}
\eeq
where $(x)^{+} = \max(x,0)$ and the quantities $\{\sigma_{k}^{(1,t)},\sigma_{c}^{(1,t)}\}$ and $\{\sigma_{k}^{(2,t)},\sigma_{c}^{(2,t)}\}$ are defined in Appendix A. The former represent the signal powers of private and common messages at iteration $t$, respectively, while the latter can be interpreted as the corresponding leakage powers. This is why \eqref{eq:optimal_k} and \eqref{eq:optimal_c} are called interference leakage-aware water-filling (ILA-WF) power allocations \cite{ThomasSPAWC2018}. Note that the Lagrange multiplier $\mu^{(t)}$ needs to satisfy the power constraint in \eqref{eq_wc_opt_max_sum_SE} and can be computed by a bisection method \cite{BoydCambridgePress2004}. The entire procedure is summarized through Algorithm 1. 
\vspace{-0.5mm}\begin{algorithm}[t]
\small
\caption{ILA-WF power allocation}\label{euclid}
\begin{algorithmic} [1]\label{euclid}
    \STATE \textbf{initialize} $t=0$ and $\rho_c^{(0)}=0$ (no RS) and $\rho_k^{(0)}= \rho_T/K$. Also,  $\mu^{(0)} = \frac{1}{2}(\mu^{(0)}_{u}+\mu^{(0)}_{l})$ with $\mu^{(0)}_{u} = 10^5$ (or some very large value) and $ \mu^{(0)}_{l} = 0$.
   
  \REPEAT	
    \FOR {$k=1$ to $K$}
      \STATE \textbf{compute} $\sigma_{k}^{(1,t)}$ and $\sigma_{k}^{(2,t)}$
            \STATE  \textbf{use} $\mu^t$ to update $p_k^t$ in \eqref{eq:optimal_k}
    \ENDFOR
    
    \STATE \textbf{compute} $\sigma_{c}^{(1,t)}$ and $\sigma_{c}^{(2,t)}$
    \STATE \textbf{use} $\mu^t$ to update $p_c^t$ in \eqref{eq:optimal_c}

\IF{$\rho_c^{(t)}+\sum\limits_k\rho_k^{(t)} > \rho_T$} \STATE {$\mu^{(t+1)}_l = \mu^{(t)},\, \mu^{(t+1)}_u = \mu_u^{(t)}$} \ELSE \STATE {$\mu^{(t+1)}_u = \mu^{t},\, \mu^{(t+1)}_l = \mu_l^{(t)}$} \ENDIF

\STATE \textbf{update} $\mu^{(t+1)} = \frac{\mu^{(t+1)}_{u}+\mu^{(t+1)}_{l}}{2}$
    \STATE \textbf{update} $t=t+1$
  \UNTIL {convergence}
  \end{algorithmic}
\vspace{-1mm}\end{algorithm}
\setlength{\textfloatsep}{0pt}
%

As done for ${\gamma}_{k,c}$ and $\gamma_{k}$, we observe that all the quantities involved in the computation of $\{\sigma_{k}^{(1)},\sigma_{c}^{(1)}\}$ and $\{\sigma_{k}^{(2)},\sigma_{c}^{(2)}\}$ are deterministic and can be computed by means of Monte Carlo simulations for any choice of the precoding scheme for the common message. Closed form expressions are provided below for a MR-inspired precoding scheme.

%
%
%
%
%
%
\vspace{-1mm}\subsection{Precoding design for common message}
 The optimal design of the precoding vector ${\bf w}_c$ for the common message requires to solve a multi-objective problem involving $\gamma_{l_{\min},c}$ and $\{{\gamma}_{i}: \forall i\}$. To overcome this issue, we assume that the difference $\mathbb{E} \{  | \vect{h}_{k}^{\Htran}  {\vect{w}}_{c} |^2 \} - | \mathbb{E}\{\vect{h}_{k}^{\Htran} {\vect{w}}_{c}  \} |^2$ in \eqref{eq:gamma_c} is small so that it can be neglected. The precoding vector is then suboptimally selected as the solution to the following problem:
 \beq
\max\limits_{\bmw_c} \min\limits_{k} \pi_k |\mathbb{E}\{ \vect{h}_{k}^{\Htran}{\vect{w}}_{c} \} |^2\quad \mbox{s.t.} \quad  \mathbb{E}\{\|  \vect{w}_{c} \|^2\}  =1
\label{eq_wc_opt_max_min}
\eeq
where 
\vspace{-1mm}\beq
\pi_k = \frac{ 1}{ \sum\limits_{i=1}^{K} \rho_i\mathbb{E} \{  | \vect{h}_{k}^{\Htran}\vect{w}_{i}   |^2 \} + \sigma^2}.
\eeq
Following \cite{MDaiTWC2016}, we heuristically select ${\bf w}_c$ as a linear combination of the estimated channel vectors $\{\bmhh_{i}: \forall i\}$:
\vspace{-1mm}\beq
\bmw_c = \alpha \sum\limits_{i=1}^K a_i \bmhh_{i}
\label{eq_wc}
\vspace{-1mm}\eeq 
where $\alpha$ is needed to satisfy the constraint $\mathbb{E}\{\|  \vect{w}_{c} \|^2\} = 1$.
Plugging \eqref{eq_wc} into $\mathbb{E}\{ \vect{h}_{k}^{\Htran}{\vect{w}}_{c} \} $, 
we may rewrite \eqref{eq_wc_opt_max_min} as:
\vspace{-1mm}\beq
\max\limits_{\{a_i\}} \min\limits_{k} \pi_k \left|\sum\limits_{i=1}^Ka_i\tr\{\vect{R}_{i} \vect{Q} ^{-1} \vect{R}_{k} \}\right|^2
\label{eq_max_min_wc_revised}
\vspace{-1mm}\eeq
where we have neglected the scaling factor $\alpha^2$. We now observe that \eqref{eq_max_min_wc_revised} can be reformulated as a geometric programming problem \cite{BoydCambridgePress2004}: 
\vspace{-1mm}\beq
\max\limits_{t>0} \; t, \,\,\, \mbox{s.t.} \,\,\,
\vect{a}^{\Ttran}\vect{u}_i \leq t, \,\, \forall\,\, i =1,..,K
\label{eq_max_min_wc_revised_GP}
\eeq
where we have defined $\vect{a} = [a_1,...,a_K]^{\Ttran}$ and $\vect{u}_i = [{u}_i(1),\ldots, {u}_i(K)]^{\Ttran}$ with entries ${u}_i(k)=\tr\{\vect{R}_i\vect{Q}^{-1}\vect{R}_k\}$. Once the solution $\vect{a}^\star$ to \eqref{eq_max_min_wc_revised_GP} is computed, the optimal $\bmw_c^\star$ is obtained as:
\vspace{-1mm}\begin{align}\label{eq:common_expected_norm}
\bmw_c^\star  =  
\frac{\sum\limits_{i=1}^K a_i^\star \bmhh_{i}}{\sqrt{\sum\limits_{i=1}^K\sum\limits_{j=1}^K a_i^\star a_j^\star \tr\{\vect{R}_i\vect{Q}^{-1}\vect{R}_j\}}}.
\vspace{-1mm}\end{align} 
The expectations that depend on $\bmw_c^\star$ can be computed in closed form as follows. By using \eqref{eq:correlated_channel_estimates} into \eqref{eq:common_expected_norm} yields
\begin{align}\label{eq:common_expected_gain}
\mathbb{E}\{ \vect{h}_{k}^{\Htran}{\vect{w}}_{c}^\star \}  = \frac{\sum\limits_{i=1}^Ka_i^\star\tr\{\vect{R}_{i} \vect{Q} ^{-1} \vect{R}_{k} \}}{\sqrt{\sum\limits_{i=1}^K\sum\limits_{j=1}^K a_i^\star a_j^\star \tr\{\vect{R}_i\vect{Q}^{-1}\vect{R}_j\}}}.
\end{align}
To compute $\mathbb{E} \{  | \vect{h}_{k}^{\Htran} \vect{w}_{c}^\star  |^2 \}$, observe that it can be rewritten as
\begin{align}\notag
&\mathbb{E} \{  | \vect{h}_{k}^{\Htran} \vect{w}_{c}^\star  |^2 \} = \frac{1}{\sum\limits_{i=1}^K\sum\limits_{j=1}^K a_i^\star a_j^\star \tr\{\vect{R}_i\vect{Q}^{-1}\vect{R}_j\}}\times \\&\!\!\!\!\!\!\left(\sum\limits_{i=1}^K{(a_i^\star)}^2 \mathbb{E} \{  |\bmhh_{i}^{\Htran} \vect{h}_{k}|^2\}+\!\sum\limits_{i=1}^K\!\sum\limits_{j=1,j\ne i}^K\!\!\!\!\!a_i^\star a_j^\star \mathbb{E} \{  \vect{h}_{k}^{\Htran} \bmhh_{i}\bmhh_{j} \vect{h}_{k}\}\!\right).\label{eq:common_expected_gain}
\end{align}
The first term in \eqref{eq:common_expected_gain} becomes (e.g., \cite[ Eq. (C.65)]{massivemimobook})
\begin{align}
\mathbb{E} \{  |\bmhh_{i}^{\Htran} \vect{h}_{k}|^2\} = \tr\{\bmR_{k}\bPhi_{i}\} + \Big|\tr\{\bmR_{k}\vect{Q}^{-1}\vect{R}_i\}\Big|^2
\end{align}
while the second one in \eqref{eq:common_expected_gain} reduces to
\begin{align}
\mathbb{E} \{  \vect{h}_{k}^{\Htran} \bmhh_{i}\bmhh_{j}^{\Htran} \vect{h}_{k}\} &\mathop{=}^{(a)} \mathbb{E} \{  \vect{h}_{k}^{\Htran} \bmhh_{i} \bmhh_{i}^{\Htran}\vect{R}_{i}^{-1}\vect{R}_{j}\vect{h}_{k}\}\\&\mathop{=}^{(b)} \tr\left\{\vect{R}_{i}^{-1}\vect{R}_{j} \mathbb{E} \{\bmhh_{k}\bmhh_{k}^{\Htran}\bmhh_{i} \bmhh_{i}^{\Htran}\}\right\}\notag\\&\quad+\tr\left\{\vect{R}_{i}^{-1}\vect{R}_{j} \mathbb{E} \{\widetilde{\vect{h}}_{k}\widetilde{\vect{h}}_{k}^{\Htran}\}\mathbb{E} \{\bmhh_{i} \bmhh_{i}^{\Htran}\}\right\}\\&\mathop{=}^{(c)} \tr\left\{\vect{R}_{i}^{-1}\vect{R}_{j} \mathbb{E} \{\bmhh_{k}\bmhh_{k}^{\Htran}\bmhh_{i} \bmhh_{i}^{\Htran}\}\right\}\notag\\&\quad+\tr\left\{\vect{R}_{i}^{-1}\vect{R}_{j} (\vect{R}_{k}-\bPhi_{k})\bPhi_{i}\}\right\}
\label{eq_hk_hhi}
\end{align}
where ${(a)}$ uses $\widehat{\vect{h}}_{j}   = \vect{R}_{j}\vect{R}_{i}^{-1}\widehat{\vect{h}}_{i}$ (as it follows from \eqref{eq:correlated_channel_estimates}), ${(b)}$ follows from $\vect{h}_{k} = \widetilde{\vect{h}}_{k} + \widehat{\vect{h}}_{k}$ and the independence between the estimate $\widehat{\vect{h}}_{k}$ and estimation error $\widetilde{\vect{h}}_{k}$, whereas ${(c)}$ uses $\mathbb{E} \{\widetilde{\vect{h}}_{k}\widetilde{\vect{h}}_{k}^{\Htran}\}\mathbb{E} \{\bmhh_{i} \bmhh_{i}^{\Htran}\} = (\vect{R}_{k}-\bPhi_{k})\bPhi_{i}$. In Appendix B, it is shown that \begin{align}\notag
\mathbb{E} \{\bmhh_{k}\bmhh_{k}^{\Htran}\bmhh_{i} \bmhh_{i}^{\Htran}\} &= \tr\{\bmB_{ik} \}\bPhi_k \\ &+ \bPhi_k^{1/2}\big(\diag(\bmB_{ik} ) + \bmB_{ik} \big){(\bPhi_k^{1/2})}^{\Htran}
\end{align}
where $\bmB_{ik} = {(\bPhi_k^{1/2})}^{\Htran}\vect{R}_{i}\vect{R}_{k}^{-1}\bPhi_k^{1/2}$ and $\diag(\cdot)$ indicates the main diagonal of the enclosed matrix.

Note that, by using the above expressions and those in \eqref{eq:MR_expectation1} and \eqref{eq:MR_expectation2}, we can eventually compute in closed form all the expectations involved in \eqref{eq:gamma_c} and \eqref{eq:gamma_p}.

\section{Simulation Results}

To quantify the  SE that can be achieved in MaMIMO with RS, we consider a cell of size $250$\,m $ \times\,  250$\,m. The UL pilot power is $\rho_{\rm{tr}}= 20$\,dBm, whereas the noise power in UL and DL is $\sigma^2 = -94$\,dBm. The samples per coherence block are $\tau = 200$ with $\tau_p = 10$. Each BS is equipped with a uniform linear array with half-wavelength antenna spacing.  Each channel consists of $S= 6$ scattering clusters, which are modeled by the Gaussian local scattering model \cite[Sec. 2.6]{massivemimobook}. Hence, the $(m_1,m_2)$th element of $\vect{R}_{i}$ is
\begin{align}\notag
&\left[ \vect{R}_{i} \right]_{m_1,m_2} =\beta_{i} \times\\&\frac{1}{S} \sum_{s=1}^S e^{\mathsf{j}\pi (m_1-m_2) \sin({\varphi}_{i,s}) }  e^{-\frac{\sigma_{\varphi}^2}{2}\left(\pi (m_1-m_2) \cos({\varphi_{i,s}}) \right)^2}\label{correllatedChannelModel}
\end{align}
where $\beta_{i}$ is the large-scale fading coefficient given by (in dB)
\begin{align}
\beta_{i}|_{\rm{dB}} =  -34.53 - 38 \, \log_{10} \left( \frac{d_{i}}{1\,\textrm{km}} \right) + F_{i}
\end{align}
with UEs being placed uniformly at random  and $d_{i} \,(>= 35 \,$m$)$ represents the distance of UE $i$ from the BS. $F_{i} \sim \mathcal{N}(0,10)$ is the logarithm of the shadow fading between UE $i$ and BS. Also, let $\varphi_{i}$ be the geographical angle to UE $i$ as seen from the BS. Cluster $s$ is characterized by the randomly generated nominal
angle-of-arrival $\varphi_{i,s}\sim \mathcal{U}[\varphi_{i}-40^\circ, \varphi_{i}+40^\circ]$ and the angles of the multipath components are Gaussian distributed around the nominal angle with standard deviation $\sigma_{\varphi}^2= 10^\circ$.

\begin{figure}[t]
 \centerline{\includegraphics[width=8cm,height=5cm]{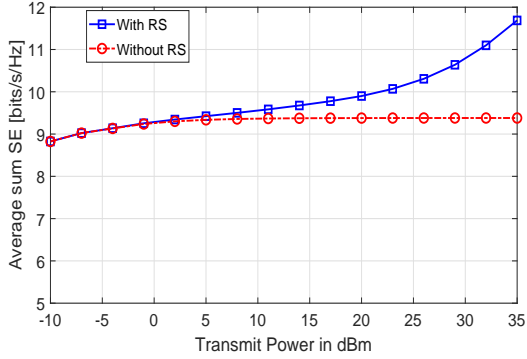}}\vspace{-0.1cm}
\caption{Sum SE versus transmit power, with $M=100$ and $K = 10$.}
\label{Fig_SE_vs_SNR_Diff_Ns}\vspace{-0cm}
\end{figure}

Fig.~\ref{Fig_SE_vs_SNR_Diff_Ns} plots the sum SE as a function of the total transmit power defined as $\rho_T$ (in dBm) with $M =100$ and $K=10$. Comparisons are made with a classical MaMIMO system with MR precoding and power allocated through Algorithm 1 with $\rho_c$ fixed to $0$. As seen, RS improves the sum SE significantly for values of $\rho_T$ higher than $5$\,dBm. Moreover, the sum SE with RS does not saturate at high $\rho_T$ values. This in contrast to what happens without RS, due to pilot contamination.
\begin{figure}[t]
 \centerline{\includegraphics[width=8cm,height=5cm]{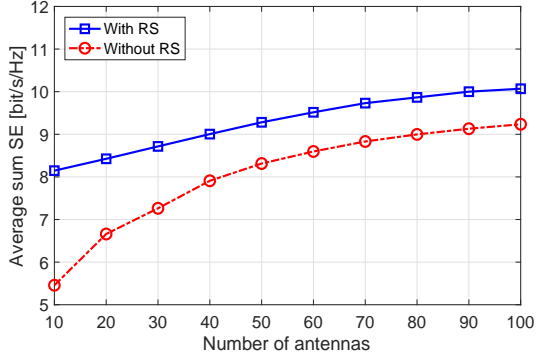}}\vspace{-0.1cm}
\caption{Sum SE versus number of antennas with $K=10$ and $\rho_T = 20$ dBm.}
\label{Fig_VaryingM}
\end{figure}

Fig.~\ref{Fig_VaryingM} illustrates the sum SE as a function of number of antennas, $M$, with $K=10$ and transmit power $\rho_T=20$ dBm. We observe that the RS scheme does help to mitigate the pilot contamination effect for a finite number of antennas.
\begin{figure}[t]
 \centerline{\includegraphics[width=8cm,height=5cm]{{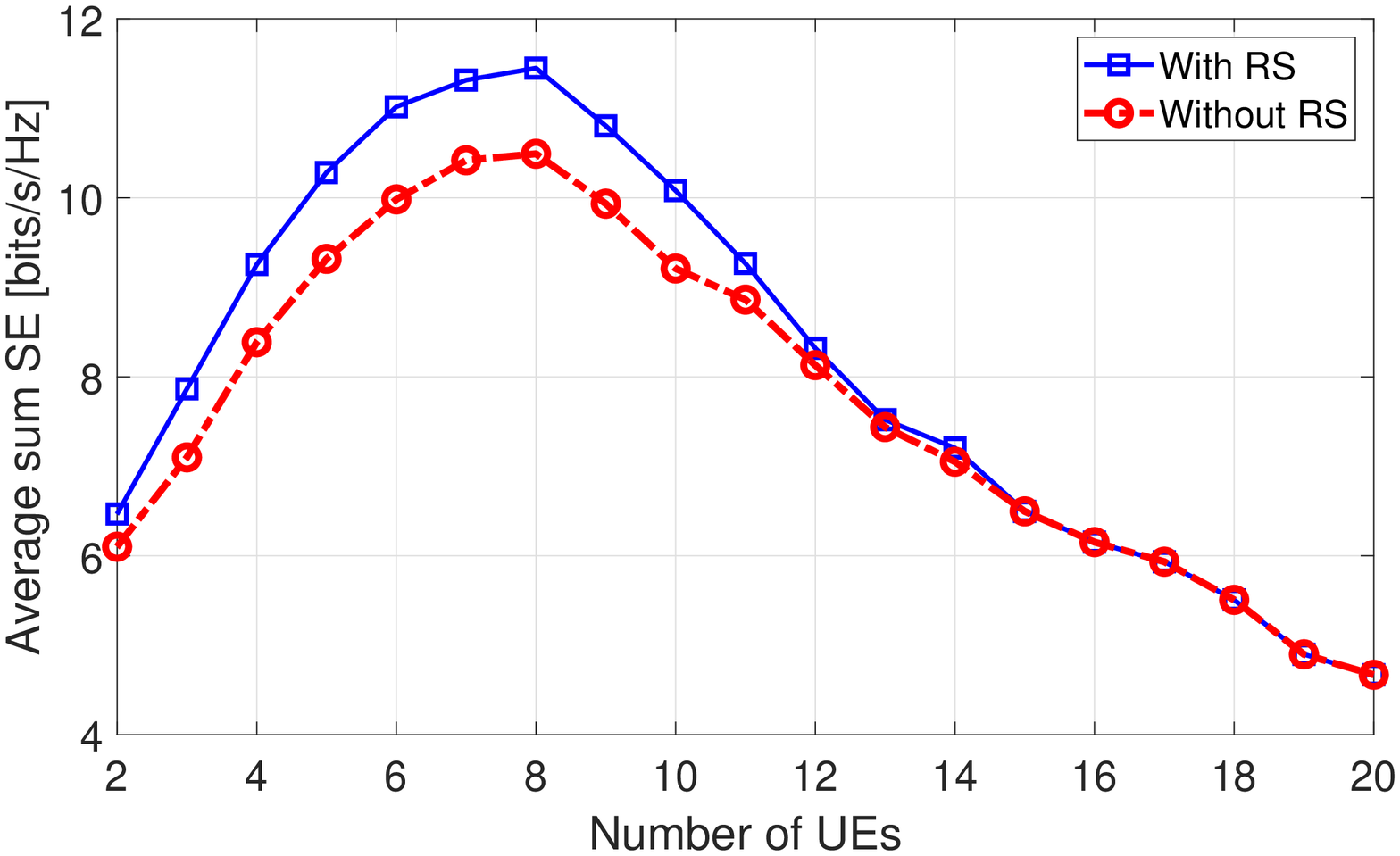}}}\vspace{-0.1cm}
\caption{Sum SE versus number of UEs with $M=100$ and $\rho_T = 20$ dBm.}
\label{Fig_VaryingK}
\end{figure}

Finally, in Fig.~\ref{Fig_VaryingK} we report the sum SE as a function of $K$ with $M=100$ and $\rho_T=20$ dBm. As $K$ increases, the gain provided by RS decreases. The larger $K$, the lower the common rate since the common message has to be decoded by all UEs. This issue can be solved by using HRS approach as in \cite{MDaiTWC2016}; this is an interesting topic left for future work.
\section{Conclusions}
This paper focused on a single-cell MaMIMO system in which all the UEs use the same pilot signal in the training phase. To deal with the reduced channel estimation quality, caused by pilot contamination, a single layer RS approach was proposed and shown to improve the SE at high SNR values. However, we remark that much remains to be done, for e.g. extension of the current work to a multi-cell setting and the design of an efficient RS message scheme to mitigate the inter-cell and intra-cell interference.
\begin{figure*}
\begin{align}\tag{45}\label{eq:sigma_1}
\sigma_{k}^{(1)} =
\frac{\mathbb{E} \{  | \vect{h}_{k}^{\Htran} \vect{w}_{k}  |^2 \}}{{\sigma^2 + \widehat{\rho}_c\Big(\mathbb{E} \{  | \vect{h}_{k}^{\Htran} \vect{w}_{c}  |^2 \} - | \mathbb{E}\{ \vect{h}_{k}^{\Htran}\vect{w}_{c}  \} |^2\Big) + \sum\limits_{i=1, i\ne k}^{K} \widehat{\rho}_i\mathbb{E} \{  | \vect{h}_{k}^{\Htran} \vect{w}_{i}  |^2 \}}}
\end{align}
\hrule
\end{figure*}
\section*{Appendix A}
Let's consider without loss of any generality the optimization of $\rho_k^{(t)}$ for given values of $\{\rho_i^{(t)}:\forall i\ne k\}$ and $\rho_c^{(t)}$. For simplicity, we drop the iteration index $t$. We begin by rewriting the SE of UE $k$ as (by explicating its dependence from $\rho_k$)
\begin{align}\notag 
\mathacr{SE}_{k}(\rho_k)  &= \frac{\tau_d}{\tau} \log_2   \left(\frac{\mathacr{NUM}_{k} (\rho_k) }{\mathacr{DEN}_{k}(\rho_k)} \right)\\&= \frac{\tau_d}{\tau} \left(\log_2   \left(\mathacr{NUM}_{k}(\rho_k) \right) -\log_2   \left(\mathacr{DEN}_{k}(\rho_k) \right)\right) \label{eq:Appendix-downlink-SE-expression-forgetbound}
\end{align}
where $\mathacr{DEN}_{k}(\rho_k)$ represents the denominator of $\gamma_k$ in \eqref{eq:gamma_p} while $\mathacr{NUM}_{k} (\rho_k)= \mathacr{DEN}_{k}(\rho_k) + {\rho_k}| \mathbb{E}\{ \vect{h}_{k}^{\Htran} \vect{w}_{k} \} |^2 $. Observe that $-\log_2   \left(\mathacr{DEN}_{k}(\rho_k)\right)$ is a non-concave function of $\rho_k$. By linearizing it around a tentative value $\widehat{\rho}_k$, the following approximation is obtained:
\begin{align}\notag
\log_2   \left(\mathacr{DEN}_{k}(\rho_k)\right) \approx \underbrace{\frac{\mathbb{E} \{  | \vect{h}_{k}^{\Htran} \vect{w}_{k}  |^2 \}-| \mathbb{E}\{\vect{h}_{k}^{\Htran} {\vect{w}}_{k}  \} |^2}{\mathacr{DEN}_{k}(\widehat{\rho}_k)}}_{\triangleq\alpha_k}(\rho_k-\widehat{\rho}_k)
\end{align}
where the terms independent of $\rho_k$ have been neglected for simplicity. Similarly, we can rewrite $\mathacr{SE}_{i}$ as
\begin{align}
\mathacr{SE}_{i}(\rho_k)  = \frac{\tau_d}{\tau} \left(\log_2   \left(\mathacr{NUM}_{i}(\rho_k) \right) -\log_2   \left(\mathacr{DEN}_{i}(\rho_k) \right)\right).
\end{align}
By linearizing both terms around $\widehat{\rho}_k$
\begin{align}
\log_2   \left(\mathacr{NUM}_{i}(\rho_k) \right) \approx \frac{\mathbb{E} \{  | \vect{h}_{i}^{\Htran} \vect{w}_{k}  |^2 \} }{\mathacr{NUM}_{i}(\widehat{\rho}_k)}(\rho_k-\widehat{\rho}_k)\\
\log_2   \left(\mathacr{DEN}_{i}(\rho_k)\right) \approx \frac{\mathbb{E} \{  | \vect{h}_{i}^{\Htran} \vect{w}_{k}  |^2 \} }{\mathacr{DEN}_{i}(\widehat{\rho}_k)}(\rho_k-\widehat{\rho}_k)
\end{align}
we obtain the following approximation for $\mathacr{SE}_{i}(\rho_k)$
\begin{align}\notag
\mathacr{SE}_{i}(\rho_k)  \approx \frac{\tau_d}{\tau}\underbrace{\left(\frac{\mathbb{E} \{  | \vect{h}_{i}^{\Htran} \vect{w}_{k}  |^2 \} }{\mathacr{NUM}_{i}(\widehat{\rho}_k)} - \frac{\mathbb{E} \{  | \vect{h}_{i}^{\Htran} \vect{w}_{k}  |^2 \} }{\mathacr{DEN}_{i}(\widehat{\rho}_k)}\right)}_{\triangleq\zeta_i}(\rho_k-\widehat{\rho}_k).
\end{align}
Following the same approach for the SE of the common message yields
\begin{align}\notag
\mathacr{SE}_{c}(\rho_k) \approx \frac{\tau_d}{\tau}\underbrace{\left(\frac{\mathbb{E} \{  | \vect{h}_{l_{\min}}^{\Htran} \vect{w}_{k}  |^2 \} }{\mathacr{NUM}_{c,\min}(\widehat{\rho}_k)} -  \frac{\mathbb{E} \{  | \vect{h}_{l_{\min}}^{\Htran} \vect{w}_{k}  |^2 \} }{\mathacr{DEN}_{c,\min}(\widehat{\rho}_k)}\right)}_{\triangleq\zeta_c}(\rho_k-\widehat{\rho}_k)
\end{align}
where $\mathacr{NUM}_{c,\min}(\rho_k)= \mathacr{DEN}_{c,\min}(\rho_k) + \rho_c| \mathbb{E}\{ \vect{h}_{l_{\min}}^{\Htran}{\vect{w}}_{c} \} |^2 $
and $\mathacr{DEN}_{c,\min}({\rho}_k)$ represents the denominator of ${\gamma}_{l_{\min},c}$ in \eqref{eq_min_gamma_c}. 
Putting all the above together, an approximation of the sum SE in \eqref{eq:sum_SE} is
\begin{align}\label{eq:SE-approximated}
\underline{\mathacr{SE}}(\rho_k) = \frac{\tau_d}{\tau} \left(\log_2   \left(\mathacr{NUM}_{k}(\rho_k)\right) - \sigma_{k}^{(2)}(\rho_k-\widehat{\rho}_k)\right)
\end{align}
where 
\beq
\sigma_{k}^{(2)} = \zeta_c + \alpha_k + \sum_{i=1,i\ne k}^K\zeta_i.
\label{eq_sigma_k2}
\eeq
 Taking the derivative of its Lagrangian (obtained after adding the power constraint in \eqref{eq_wc_opt_max_sum_SE}) and equating it to zero yields
\beq
\frac{\mathbb{E} \{  | \vect{h}_{k}^{\Htran} \vect{w}_{k}  |^2 \}}{\mathacr{NUM}_{k}(\rho_k)} -\sigma_{k}^{(2)}-\mu = 0
\eeq
from which one obtain \eqref{eq:optimal_k} in the text, with $\sigma_{k}^{(1)}$ given in \eqref{eq:sigma_1}. A similar approach for $\rho_c$ yields
\setcounter{equation}{45}
\beq\label{eq:optimal_c_app}
\frac{\mathbb{E} \{  | \vect{h}_{l_{\min}}^{\Htran} \vect{w}_{c}  |^2 \}}{\mathacr{NUM}_{c,\min}(\rho_c)} -\sigma_{c}^{(2)}-\mu = 0
\eeq
where $\sigma_{c}^{(2)}$ can be obtained as done for $\sigma_{k}^{(2)}$ in \eqref{eq_sigma_k2}; details are omitted for space limitation. Solving \eqref{eq:optimal_c_app} yields \eqref{eq:optimal_c} in the text, where $\sigma_{c}^{(1)}$ is
\begin{align}\label{eq:sigma_c}
\sigma_{c}^{(1)} =
\frac{\mathbb{E} \{  | \vect{h}_{l_{\min}}^{\Htran} \vect{w}_{c}  |^2 \}}{{\sigma^2 + \sum\limits_{i=1}^{K} \widehat{\rho}_i\mathbb{E} \{  | \vect{h}_{l_{\min}}^{\Htran} \vect{w}_{i}  |^2 \}}}.
\end{align}
\section*{Appendix B}
Rewrite $\bmhh_k = \bPhi_k^{1/2}\bmc$, with $\bmc \sim \mathcal{CN}({\bf 0}, \bmI)$ and define the deterministic matrix $\bmB_{ik} = {(\bPhi_k^{1/2})}^{\Htran}\vect{R}_{i}\vect{R}_{k}^{-1}\bPhi_k^{1/2}$. By recalling that $\widehat{\vect{h}}_{i}   = \vect{R}_{i}\vect{R}_{k}^{-1}\widehat{\vect{h}}_{k}$ yields
\begin{align}
\mathbb{E} \{\bmhh_{k}\bmhh_{k}^{\Htran}\bmhh_{i} \bmhh_{i}^{\Htran}\} = \bPhi_k^{1/2}\mathbb{E} \{\bmc\bmc^{\Htran}\bmB_{ik}\bmc\bmc^{\Htran}\}{(\bPhi_k^{1/2})}^{\Htran}.
\end{align}
It then follows that 
\begin{align}\label{eq:1_Appendix}
\mathbb{E} \left\{[\bmc\bmc^{\Htran}\bmB_{ik}\bmc\bmc^{\Htran}]_{mn}\right\} = \sum_{j=1}^M\sum_{l=1}^M[\bmB_{ik}]_{lj}\mathbb{E} \left\{ c_mc_l^*c_jc_n^*\right\}.
\end{align}
If $m=n$, then \eqref{eq:1_Appendix} is always zero except for $l=j$:
\begin{align}
\sum_{l=1}^M[\bmB_{ik}]_{ll}\mathbb{E} \left\{ |c_m|^2|c_l|^2\right\} = 3[\bmB_{ik}]_{mm} +\!\!\!\sum_{l=1,l\ne m}^M[\bmB_{ik}]_{ll}
\end{align}
where we have taken into that $\mathbb{E} \left\{ |c_m|^4\right\} = 3$.
If $m\ne n$, then \eqref{eq:1_Appendix} is always zero except for $l=m$ and $j=n$
\begin{align}
[\bmB_{ik}]_{mn}\mathbb{E} \left\{ c_mc_m^*c_mc_n^*\right\} = [\bmB_{ik}]_{mn}.
\end{align}
Putting the above results together yields $\mathbb{E} \{\bmhh_{k}\bmhh_{k}^{\Htran}\bmhh_{i} \bmhh_{i}^{\Htran}\} = \tr\{\bmB_{ik}\}\bmI + \diag(\bmB_{ik}) + \bmB_{ik}$.

\bibliographystyle{IEEEtran}
\bibliography{HBF_ref,refs}

\end{document}

%% file: defines2.tex
\newtheorem{lemma}{Lemma}

\newcommand{\beq}{\begin{equation}}
\newcommand{\eeq}{\end{equation}}

\newcommand{\diag}{\,\mbox{diag}\,}

\newcommand{\E}{\,\mbox{E}}

\def\adots{\mathinner{\mskip0mu\raise0pt\vbox{\kern7pt\hbox{.}}\mskip3mu
          \raise4pt\hbox{.}\mskip3mu\raise8pt\hbox{.}\mskip0mu}}

\newfont{\bb}{msbm10 scaled 1100}

\newcommand{\tr}{\mbox{tr}}

\newcommand{\bmh}{\mbox{\boldmath $h$}}

\newcommand{\CN}{{\cal CN}}

\newcommand{\bmw}{{\mathbf w}}

\newcommand{\bmc}{{\mathbf c}}

\newcommand{\bmQ}{{\mathbf Q}}
\newcommand{\bmR}{{\mathbf R}}

\renewcommand{\bmh}{{\mathbf h}}

\newcommand{\bmB}{{\mathbf B}}

\newcommand{\bmI}{{\mathbf I}}